\documentclass[aps,prl,twocolumn,lengthcheck]{revtex4}
\usepackage{amsfonts, amssymb, amsmath,graphicx,comment}
\usepackage{subfigure}

\begin{document}

\title{Stability of ultracold atomic Bose condensates with Rashba spin-orbit coupling against quantum and thermal fluctuations}

\author{Tomoki Ozawa}
\author{Gordon Baym}
\affiliation{
Department of Physics, University of Illinois at Urbana-Champaign, 1110 W. Green St., Urbana, Illinois 61801, USA
}

\date{\today}

\def\del{\partial}
\def\p{\prime}
\def\simge{\mathrel{%
         \rlap{\raise 0.511ex \hbox{$>$}}{\lower 0.511ex \hbox{$\sim$}}}}
\def\simle{\mathrel{
         \rlap{\raise 0.511ex \hbox{$<$}}{\lower 0.511ex \hbox{$\sim$}}}}
\newcommand{\feynslash}[1]{{#1\kern-.5em /}}

\begin{abstract}
We study the stability of Bose condensates with Rashba-Dresselhaus spin-orbit coupling in three dimensions against quantum and thermal fluctuations.
The ground state depletion of the plane-wave condensate due to quantum fluctuations is, as we show, finite, and therefore the condensate is stable. We also calculate the corresponding shift of the ground state energy. 
Although the system cannot  condense in the absence of interparticle interactions, we show by estimating the number of excited particles that interactions stabilize the condensate even at non-zero temperature.  Unlike in the usual Bose gas, the normal phase is not kinematically forbidden at any temperature; calculating the free energy of the normal phase at finite temperature, and comparing with the free energy of the condensed state, we infer that generally the system is condensed at zero temperature, and undergoes a transition to normal at non-zero temperature.
\end{abstract}

\maketitle

    Creation of artificial gauge fields in ultracold atoms presents opportunities to study physical phenomena previously unattainable in atomic or condensed matter systems \cite{Dalibard2011}.
Of current interest is the non-Abelian Rashba-Dresselhaus spin-orbit coupling \cite{Rashba1960,Dresselhaus1955} which, following several proposals \cite{Liu2009,Juzeliunas2010,Campbell2011,Anderson2011},  has recently been realized experimentally  \cite{Lin2011,Williams2012}.   Depending on the details of the  Rashba-Dresselhaus coupling and interparticle interactions,
one of two possible ground state phases is expected: a plane-wave state, in which the system condenses into a single momentum state, and a striped phase, where the condensate is a linear superposition of two states with opposite momenta [10, 11, reviewed in 12]; renormalization of the interparticle interaction beyond mean-field \cite{Ozawa2011,Gopalakrishnan2011} has significant effects on the phase diagram \cite{Gopalakrishnan2011,Ozawa2012}.

    Condensates of ultracold bosons in three dimensions with Rashba spin-orbit coupling differ from usual Bose-Einstein condensates (BEC's)  in several important ways.  In the absence of interparticle interactions, the low-lying density of states is two-dimensional  \cite{Stanescu2008}, and thus condensation is destroyed by thermal fluctuations at any non-zero temperature.  With interparticle interactions present, fluctuations around mean-field states lead at finite temperature to an instability of the plane-wave state in two dimensions \cite{Jian2011}.  In this paper, we focus on three-dimensional ultracold bosons with isotropic Rashba-Dresselhaus coupling in the x-y plane, to investigate the effects of quantum and thermal fluctuations on a plane-wave Bose-Einstein condensation, and show that interactions in fact stabilize the condensate in 3D.
This interaction-induced BEC is a unique feature of bosons with Rashba-Dresselhaus spin-orbit coupling, with no analogous system yet found.  However, unlike in usual BEC's, a non-condensed state is not, as we show, kinematically forbidden at any non-zero temperature.  Condensation, while  favored at very low temperature, should disappear at high temperature.  As in a BCS superconductor, where both a normal and condensed state are allowed at low temperature, the system should undergo a similar phase transition at a critical temperature.  

We consider bosons with an isotropic Rashba-Dresselhaus spin-orbit coupling in three dimensions with an isotropic interaction, described by the Hamiltonian
\begin{align}
	\mathcal{H}
	&=
	\sum_{\mathbf{p}}
	\begin{pmatrix}
	a_\mathbf{p}^\dagger & b_\mathbf{p}^\dagger
	\end{pmatrix}
	\left[
	\frac{p^2 + \kappa^2}{2m}I + \frac{\kappa}{m} (\sigma_x p_x + \sigma_y p_y)
	\right]
	\begin{pmatrix}
	a_\mathbf{p} \\
	b_\mathbf{p}
	\end{pmatrix}
	\notag \\
	&+
	\frac{g}{2V}
	\sum_{\mathbf{p}_1 + \mathbf{p}_2 = \mathbf{p}_3 + \mathbf{p}_4}
	\left(
	a_{\mathbf{p}_4}^\dagger a_{\mathbf{p}_3}^\dagger a_{\mathbf{p}_2} a_{\mathbf{p}_1}
	\right.
	\notag \\
	&\hspace{1cm}\left.
	+
	b_{\mathbf{p}_4}^\dagger b_{\mathbf{p}_3}^\dagger b_{\mathbf{p}_2} b_{\mathbf{p}_1}
	+
	2 a_{\mathbf{p}_4}^\dagger b_{\mathbf{p}_3}^\dagger b_{\mathbf{p}_2} a_{\mathbf{p}_1}
	\right).
\end{align}
As previously \cite{Ozawa2011,Ozawa2012}, 
$m$ is the atomic mass, $V$ is the volume of the system, and $\kappa$ is the spin-orbit coupling strength, taken to be positive.
The isotropic s-wave coupling is $g$. The operators $a_\mathbf{p}$ and $b_\mathbf{p}$ annihilate atoms with momentum $\mathbf{p}$ in the pseudospin states $a$ and $b$, respectively.
The $\sigma_x$ and $\sigma_y$ are the usual Pauli matrices between the internal states, and 
$I$ is the two-by-two identity matrix.
The dispersion relation of the single particle terms in the Hamitonian has two branches
$\epsilon_\pm (\mathbf{p}) = \{(p_\perp \pm \kappa)^2 + p_z^2\}/2m$, where $p_\perp \equiv \sqrt{p_x^2+p_y^2}$,
with circularly degenerate ground states along $(p_\perp,p_z) = (\kappa, 0)$.
In this paper, we assume that $g$ is the (constant) mean-field coupling; extension beyond mean-field coupling, as in \cite{Gopalakrishnan2011,Ozawa2011,Ozawa2012}, is left as a future problem.
Starting from the plane-wave ground state with momentum $\boldsymbol\kappa \equiv (\kappa, 0, 0)$, we construct the single-partice Green's functions, including quantum fluctuations via the Bogoliubov approximation, and derive the low-momentum spectra, estimate the condensate depletion, and calculate the ground state energy as a function of the spin-orbit coupling strength.  As we show, the number of excited particles increases and the ground state energy decreases with increasing spin-orbit coupling strength.  

Since the operator $(a^\dagger_{\boldsymbol\kappa} - b^\dagger_{\boldsymbol\kappa})/\sqrt{2}$ creates a particle in the single-particle ground state with momentum $\boldsymbol\kappa$,  it is easier to work in the following $(-,+)$ basis:
\begin{align}
	\begin{pmatrix}
	\psi_{-,\mathbf{p}} \\ \psi_{+,\mathbf{p}}
	\end{pmatrix}
	\equiv
	\frac{1}{\sqrt{2}}
	\begin{pmatrix}
	1 & -1 \\ 1 & 1
	\end{pmatrix}
	\begin{pmatrix}
	a_\mathbf{p} \\ b_\mathbf{p}
	\end{pmatrix}.
\end{align}
The state created by $\psi^\dagger_{-,\boldsymbol\kappa}$ is macroscopically occupied.

We first derive the fluctuations of the system in terms of the single particle matrix Green's functions
with anomalous components, 
\begin{align}
	\mathbf{G}(\mathbf{q},t_1-t_2)
	\equiv
	-i \langle T\left( \Psi_\mathbf{q}(t_1) \Psi^\dagger_\mathbf{q}(t_2)\right) \rangle,
\end{align}
where the four-component spinor $\Psi_\mathbf{q}(t)$ is
\begin{align}
	\Psi_\mathbf{q}(t)
	&\equiv
	\left(
	\psi_{-,\boldsymbol\kappa + \mathbf{q}}(t), \psi^\dagger_{-,\boldsymbol\kappa-\mathbf{q}}(t),
	\right.
	\notag \\
	&\hspace{2cm}
	\left.
	\psi_{+,\boldsymbol\kappa+\mathbf{q}}(t), \psi^\dagger_{+,\boldsymbol\kappa-\mathbf{q}}(t)
	\right).
\end{align}

From the equations of motion for the Green's functions with Hamiltioninan $\mathcal{H} - \mu \mathcal{N}$,
in the Bogliubov approximation, where the operators $\psi^\dagger_{-,\boldsymbol\kappa}$ and $\psi_{-,\boldsymbol\kappa}$ are replaced by $\sqrt{N_0}$ with $N_0$ the number of condensate particles, and with the Hartree-Fock energy included, we obtain 
\begin{align}
	&\mathbf{G}^{-1}(\mathbf{q},z) =
	\begin{pmatrix}
	z - A & -gn_0 & i\frac{\kappa}{m}q_y & 0
	\\
	-gn_0 & -z - A & 0 & i\frac{\kappa}{m}q_y
	\\
	-i\frac{\kappa}{m}q_y & 0 & z - B & 0
	\\
	0 & -i\frac{\kappa}{m}q_y & 0 & -z - D
	\end{pmatrix}, \label{greenmotion}
\end{align}
where $\mathbf{G}(\mathbf{q}.z)$ is the Fourier transform of $\mathbf{G}(\mathbf{q},t)$
and
\begin{align}
	A(\mathbf{q}) & \equiv q^2/2m -\mu +g( 2n_0+2n_-+n_+) 
	\notag \\
	B(\mathbf{q}) & \equiv (2\boldsymbol\kappa + \mathbf{q})^2 / 2m -\mu +g( n_0+n_-+2n_+) \label{bdef}
	\notag \\
	D(\mathbf{q}) & \equiv B(-\mathbf{q}).
\end{align}
The chemical potential in leading order is
$\mu_0 = \partial \langle \mathcal{H} \rangle/\partial N_0 = gn_0 + 2gn_- + gn_+$,
where $n_0 = N_0/V$, and
\begin{align}
	n_\mp = \frac{1}{V}\sum_{\mathbf{p} \neq \boldsymbol\kappa} \langle \psi^\dagger_{\mp,\mathbf{p}}\psi_{\mp,\mathbf{p}} \rangle
\end{align}
are the number of particles in the $(-)$ and $(+)$ states that are {\it not} in the condensate.
In lowest order,
\begin{align}
	A(\mathbf{q}) & = q^2/2m + gn_0,
	&
	B(\mathbf{q}) & = (2\boldsymbol\kappa + \mathbf{q})^2 / 2m. \label{ablowest}
\end{align}

{\it Low-momentum excitations:}
The excitation spectra is given by the poles of $\mathbf{G}(\mathbf{q}.z)$ with $\mu = \mu_0$;
the poles satisfy
\begin{align}
	&0 = \det \mathbf{G}^{-1}(\mathbf{q},z)
	= (gn_0)^2 (z-B)(z+D) + 
	\notag \\
	&
	\left[(z-A)(z-B) - \tfrac{\kappa^2}{m^2} q_y^2 \right]
	\left[(z+A)(z+D) - \tfrac{\kappa^2}{m^2} q_y^{2} \right].
\end{align}
Since $\det \mathbf{G}^{-1}(\mathbf{q},z) = \det \mathbf{G}^{-1}(-\mathbf{q},-z)$, the roots come in pairs: two positive and two negative, corresponding to two excitations, for each $\mathbf{q}$, 

One of the two excitations
is gapless in the limit $\mathbf{q} \to 0$, and the other is gapless in the limit $\mathbf{q} \to -2\boldsymbol\kappa$.
Although the roots of $\det \mathbf{G}^{-1}(\mathbf{q},z) = 0$ cannot be found analytically for general $\mathbf{q}$,
we can construct the low energy dispersion relations for
momenta around the gapless points, $|\mathbf{q}| \ll \kappa$ and $|\mathbf{q} + 2\boldsymbol\kappa| \ll \kappa$.
With strong spin-orbit coupling, $\kappa^2 / m \gg g |n_+ - n_-| \equiv g|\Delta n|$, the spectrum to leading order in the low momentum limit $|\mathbf{q}| \ll \kappa$ is
\begin{align}
	&
	\epsilon_1(\mathbf{q})
	\approx
	\sqrt{2gn_0}
	\left[
	\frac{q_x^2+q_z^2}{2m} + \frac{q_y^2}{4\kappa^2}\left( g \Delta n + \frac{q_y^2}{2m}\right)
	\right]^{1/2} 	\label{lowdis}
\end{align}
The dispersion relation for $q_y=0$ is linear at low momenta, as in the usual Bogoliubov spectrum \cite{Wu2011}.  Since
 $q_y^2/2m$ is generally larger than  $g|\Delta n|$  in typical experimental setups, the dispersion is essentially quadratic for $q_x = q_z =0$ \footnote{As discussed below, $\Delta n$ is of order $n_0 \sqrt{(mg)^3 n_0}$, and $n_0 \sim N/L^3$, where $N$ is the total number of particles, and $L$ is the linear size of the system,  and the smallest
$q_y$ is $\sim \pi /L$.  Then naively writing $g \sim 4\pi a/m$ where $a$ is the scattering length, we obtain  $|g\Delta n|/ q_y^2/2m \sim10^2 N^{3/2}(a/L)^{5/2}$.
Taking typical experimental parameters from \cite{Lin2011}, $N \sim 10^5 $, $L \sim 10^{-2}$cm, and $a \sim $5nm we estimate  $|g\Delta n|< 0.1 q_y^2/2m $.}. 
On the other hand, the gapless spectrum in the limit of small $\mathbf{q}^\prime \equiv \mathbf{q} + 2\boldsymbol\kappa$ is
\begin{align}
	\epsilon_2 (\mathbf{q}^\prime)
	=
	\frac{q_x^{\prime 2} + q_z^{\prime 2}}{2m} + \frac{gn_0}{\kappa^2/m + gn_0}\frac{q_y^{\prime 2}}{4m}.
	\label{lowdis2}
\end{align}
This excitation is quadratic and free-particle like.
The spectra of the two excitations agree with the result of \cite{Barnett2012}.

{\it Condensate depletion:}
The densities $n_\pm$ are given in terms of the Green's functions by
\begin{align}
	n_-
	&=
	i \int \frac{dz d^3 q}{(2\pi)^4} G_{11}(\mathbf{q},z)  \notag
	\\
	n_+
	&=
	i\int \frac{dz d^3 q}{(2\pi)^4} G_{33}(\mathbf{q},z), \label{npgreen}
\end{align}
where the contour in the $z$ integration surrounds the negative poles in the positive sense.  To a first approximation we neglect $n_\mp$ inside $\mathbf{G}(\mathbf{q},z)$ .  
We assume that $\sqrt{(2mg)^3 n_0}$ is small, and in the end we see that
 $\Delta n/n_0$ is small, justifying ignoring $n_\pm$ in the integration.

Before evaluating the integrals in (\ref{npgreen}), we show that they converge in the ultraviolet. Explicitly,\begin{align}
	n_-
	=&
	i\int \frac{dz d^3 q}{(2\pi)^4} \frac{(z-B)}{\det \mathbf{G}^{-1}}
	\left[(z+A)(z+D) - \frac{\kappa^2 q_y^2}{m^2} \right].
	\notag \\
	\label{nmexp}
\end{align}
The poles in $z$ at large $q$ behave as $-q^2/2m$, and since at large $q$, $A \sim B \sim D \sim q^2/2m$ naive power counting would indicate that the integrand after the $z$ integration behaves as $q^{-1}$, which combined with three $q$-integrals yields a quadratic ultraviolet divergence.  In fact, cancellations in the integrand lead to convergence.  In evaluating the $z$ integral in terms of the poles of the integrand, we employ the relation derived from $\det\mathbf{G}^{-1}(\mathbf{q},z) = 0$ at the poles,
\begin{align}
	(z+A)(z+D) - \frac{\kappa^2}{m^2} q_y^{2}
	=
	-\frac{(gn_0)^2 (z-B)(z+D)}{(z-A)(z-B) - \frac{\kappa^2}{m^2} q_y^2}
	\notag \\ \label{relpoles}
\end{align}
to replace the left side of (\ref{relpoles})  in the numerator of (\ref{nmexp}) with the term on the right side.  Then 
\begin{align}
	n_-
	&=
	-i\int \frac{dz d^3 q}{(2\pi)^4}  \frac{(gn_0)^2}{\det \mathbf{G}^{-1}}\frac{ (z-B)^2(z+D)}{(z-A)(z-B) - \frac{\kappa^2}{m^2} q_y^2}. \notag \\ \label{nmconv}
\end{align}
In this form the integrand behaves explicitly as $q^{-4}$ after the $z$-integral, and is thus ultraviolet convergent;
the integral for $n_+$ is similarly convergent.

\begin{figure}[htbp]
\begin{center}
\includegraphics[width=7.0cm]{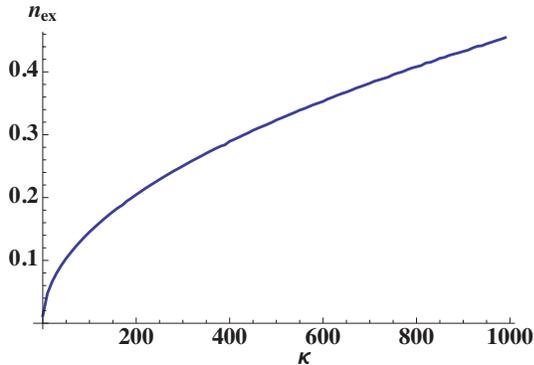}
\caption{The number of excited particles, in units of $(2mgn_0)^{3/2}$ as a function of the spin-orbit coupling strength $\kappa$ in units of $\sqrt{2mgn_0}$. Generally $n_- \gg n_+$ and $n_{ex} \approx n_-$.}
\label{nexfig}
\end{center}
\end{figure}

The depletions $n_\mp$ can be evaluated numerically as a function of $\kappa / \sqrt{2mgn_0}$.
The number of excited particles $n_{ex} = n_- + n_+$ is plotted in Fig. \ref{nexfig}. 
Generally, $n_- \gg n_+$, and the contribution of $n_+$ to the number of excited particles is negligible.
As the figure shows, the condensate depletion increases with $\kappa / \sqrt{2mgn_0}$, and is
of order $n_0 \sqrt{(2mg)^3 n_0} \ll n$, thus justifying our use of the Bogoliubov approximation.

{\it Ground state energy:}
The ground state energy density, $E$, in terms of the Green's function is \cite{Fetter}
\begin{align}
	&E = \mu n/2
	\notag \\
	&
	+
	\frac{i}{2}\int \frac{dz d^3 q}{(2\pi)^4} 
	\mathrm{Tr}\left[
	\left\{
	\left(z + \frac{(\boldsymbol\kappa+\mathbf{q})^2 + \kappa^2}{2m}\right) I
	+
	\right. \right.
	\notag \\
	&
	\left.\left.
	\frac{\kappa}{m}\left( -(\kappa+q_x) \sigma_z + q_y \sigma_y \right)
	\right\}
	\begin{pmatrix}
	G_{11}(\mathbf{q},z) & G_{13}(\mathbf{q},z) \\
	G_{31}(\mathbf{q},z) & G_{33}(\mathbf{q},z)
	\end{pmatrix}
	\right],
\end{align}
where the term in braces is $z$ plus the single-particle Hamiltonian in the $(-,+)$ basis.
The integral equals $gn_0 (2mgn_0)^{3/2}$ times a dimensionless function $X$ of $\mu / gn_0$ and $\kappa / \sqrt{2mgn_0}$.
Since the chemical potential in mean-field is $\mu = gn_0$
and $n_{ex}$ is $\mathcal{O} ((2mg n_0)^{3/2})$, the energy density is  $\mu n/2 + X gn (2mgn)^{3/2}$.
Then, writing the chemical potential similarly as $\mu = gn ( 1 + Y \sqrt{(2mg)^3n})$, and using
$\mu = \del E /\del n$, one finds $Y = -10X$; thus the ground state energy is
\begin{align}
	E \approx \frac{gn^2}{2}\left( 1 - 8X \sqrt{(2mg)^3 n}\right).
\end{align}
In calculating $X$, we take $\mu = gn_0$; deviations of $\mu$ from $gn_0$ result in higher order corrections.
For $\kappa \to 0$, one finds $X = -1/15\sqrt{2}\pi^2$, which leads to the ground state energy derived by Lee and Yang \cite{LeeYang,LeeHuangYang}.
For general $\kappa$,  we calculate $X$ numerically.

\begin{figure}[htbp]
\begin{center}
\includegraphics[width=7.5cm]{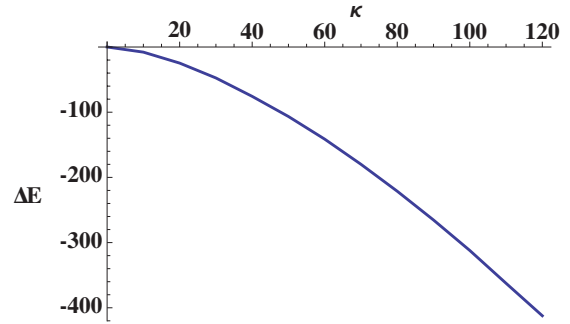}
\caption{The shift in the ground state energy density, $\Delta E$, in units of $(\sqrt{(2mg)^3 n})gn^2 / 2$, as a function of the spin-orbit coupling strength $\kappa$ in units of $\sqrt{2mgn_0}$.}
\label{energy_1}
\end{center}
\end{figure}

Figure~\ref{energy_1} shows the shift in the ground state energy, $\Delta E \equiv E - gn^2/2$, in units of $(\sqrt{(2mg)^3 n})gn^2/2$, as a function of
 $\kappa / \sqrt{2mgn_0}$.
The energy decreases with increasing $\kappa$, and $\Delta E$ changes from positive to negative at $\kappa  \sim 0.6 \sqrt{2mgn_0}$, an effect too small to see in the figure.

{\it BEC at finite temperature:}
In the absence of interparticle interactions, bosons with an isotropic Rashba spin-orbit coupling do not condense at non-zero temperature because the density of states becomes two-dimensional ($m\kappa/2\pi$) at low energy \cite{Stanescu2008}.
However, in the presence of interactions it is possible for the system to Bose condense at finite temperature, as we now discuss.

The number of excited particles at temperature $T$ is
\begin{align}
	n_{ex}
	=
	-T \sum_\nu \int \frac{d^3 q}{(2\pi)^3} \left( G_{11}(\mathbf{q},z_\nu) + G_{33}(\mathbf{q},z_\nu)\right),
\end{align}
where the $\nu$ sum is over bosonic Matsubara frequencies.
The system forms a BEC at a given temperature when $n_{ex}$ converges in the infrared, and the total particle density exceeds $n_{ex}$.  The infrared structure is captured by the $z_\nu = 0$ component of the Matsubara sum.
Since there are two gapless excitations $\epsilon_1 (\mathbf{q})$ and $\epsilon_2 (\mathbf{q}^\prime)$, we need to add infrared contributions from two limits $\mathbf{q} \to 0$ and $\mathbf{q}^\prime \to 0$.
In the limit of small $q$ and $q^\prime$, one finds from inverting Eq.~(\ref{greenmotion}), 
\begin{align}
	&G_{11}(\mathbf{q},0) + G_{33}(\mathbf{q},0)
	=
	-\frac{gn_0}{\epsilon_1 (\mathbf{q})^2},
	\notag \\
	&G_{11}(\mathbf{q}^\prime,0) + G_{33}(\mathbf{q}^\prime,0)
	=
	-\frac{1}{\epsilon_2 (\mathbf{q}^\prime)},
\end{align}
respectively, and thus
\begin{align}
	n_{ex}(\mu_0)
	\sim
	T \int \frac{d^3 q}{(2\pi)^3} \left( \frac{gn_0}{\epsilon_1(\mathbf{q})^2} + \frac{1}{\epsilon_2(\mathbf{q})} + C \right),
	\label{nexinfra}
\end{align}
where $C$ is a constant as $\mathbf{q} \to 0$.
The integral converges in the infrared, and thus a BEC can form at finite temperature.

This result is consistent with Jian and Zhai's effective field theory approach to calculate phase fluctuations \cite{Jian2011}, applied in three dimensions, through the direct relation between the condensate depletion and the phase fluctuations\cite{Baym2004}:
\begin{align}
n_0 \sim n\,e^{-\langle(\phi(r)-\phi(r'))^2\rangle/2}, \quad\quad
|r-r'|\to \infty.
\label{phasefluct}
\end{align}

{\it Normal state:}
So far, we have assumed the existence of condensate, and proved that the condensate is not destroyed by thermal fluctuations.
We should  also ask whether a non-condensed state is favorable at finite temperature.
Here we obtain the free energy of the normal state within the Hartree-Fock approximation and compare the free energies with and without a condensate.

The reduced Hamiltonian within the Hartree-Fock approximation with no condensate is
\begin{align}
	&\mathcal{H}_{\mathrm{HF}}
	=
	-Vg \left( n_-^2 + n_+^2 + n_- n_+ \right)
	\notag \\
	&+
	\sum_{\mathbf{p}}
	\begin{pmatrix}
	\psi^\dagger_{-,\mathbf{p}} & \psi^\dagger_{+,\mathbf{p}}
	\end{pmatrix}
	\begin{pmatrix}
	A & -i\frac{\kappa}{m}p_y \\
	i\frac{\kappa}{m}p_y & B
	\end{pmatrix}
	\begin{pmatrix}
	\psi_{-,\mathbf{p}} \\ \psi_{+,\mathbf{p}}
	\end{pmatrix},
\end{align}
where $A$ and $B$ are as in (\ref{bdef}) with $n_0 = 0$.
In fact, $n_- = n_+ = n/2$, where $n$ is the total number density of particles; namely there is no spontaneous imbalance of population in each pseudospin species, as one can prove by introducing independent chemical potentials for each species, and seeing,  as in Appendix B of \cite{Ozawa2010}, that the second derivative of the Ginzburg-Landau free energy with respect to the population imbalance is positive.

With $n_\pm = n/2$, the Helmholtz free energy density is
\begin{align}
	&F
	=
	\mu n - \frac{3}{4}gn^2 +
	\frac{1}{\beta V} \sum_{\mathbf{p}}
	\left\{\ln \left( 1 - e^{-\beta \xi_- (\mathbf{p})}\right)
	\right.
	\notag \\
	&\hspace{3cm}+
	\left.\ln \left( 1 - e^{-\beta \xi_+ (\mathbf{p})} \right) \right\},
\end{align}
where $\xi_\pm (\mathbf{p}) \equiv \{(p_\perp \pm \kappa)^2 + p_z^2 \}/(2m) - \mu + 3gn/2$;
the chemical potential is determined by the number equation $n = (1/V)\sum_\mathbf{p} \left\{ f (\xi_- (\mathbf{p})) + f (\xi_+ (\mathbf{p})) \right\}$, where $f (x) \equiv 1/(e^{\beta x} - 1)$.
Unlike for free bosons in three dimensions, it is possible at any temperature to find a value of $\mu$ which satisfies the number equation, thus the state without condensate is not kinematically forbidden at any non-zero temperature.

As $T \to 0$ in the absence of a condensate, $\mu \to 3gn/2$, and $F
\to 3gn^2/4$.
This energy is larger than the ground state energy with condensate,
$(gn^2 / 2) (1 + \mathcal{O}(\sqrt{(2mg)^3 n}))$.
Therefore, at sufficiently low temperature, a condensate is
energetically preferred.
At low temperature $F(\mu, n_0) < F(\mu,0)$, so $n_0>0$.
The condensate density decreases with temperature, and the transition
to the normal state, if second order, occurs when $\partial F(\mu,
n_0)/\partial n_0 =
0$ at $n_0=0$. Determination of the order of the transition, the
transition temperature, and possible
critical exponents at the transition is in progress \footnote{At
the mean field level, the transition is (spuriously) first order, as
in the Bogoliubov approximation to the finite temperature Bose gas
\cite{Markus2003}.
}.

\begin{acknowledgements}
We are grateful to Jason Ho for asking about condensate depletion. We thank Ryan Barnett for a clarifying discussion of Ref.~\cite{Barnett2012}., and T.O. thanks Philip Powell for significant help with the numerics. This research was supported in part by NSF Grant PHY09-69790.
\end{acknowledgements}

\end{document}